\documentclass[a4paper,11pt]{article}
\usepackage{pos}
\def\logo{}
\def\posurl{}

\title{Searches for heavy neutral leptons with machine learning at the CMS experiment}

\author*{Joscha Knolle}
\onbehalf{on behalf of the CMS Collaboration}
\affiliation{Universiteit Gent, Ghent, Belgium}
\emailAdd{joscha.knolle@cern.ch}

\abstract{%
    Two recent searches for heavy neutral leptons (HNLs) performed with proton-proton collision data recorded at 13\TeV by the CMS experiment are presented.
    A prompt search in the trilepton final state analyses events with exactly three charged leptons originating from the primary proton-proton interaction vertex, targeting HNL masses between 10\GeV and 1.5\TeV.
    A displaced search in the dilepton final state analyses events with exactly one prompt charged lepton and a second nonprompt charged lepton associated with a jet and a secondary vertex, targeting HNL masses between 1 and 20\GeV.
    In both searches, machine-learning methods are applied to separate the HNL signal from the standard model background.
    Exclusion limits are set on the HNL coupling strength as a function of the HNL mass, covering different mass ranges and HNL scenarios.
    In several cases, the results exceed previous limits.
}

\FullConference{%
    12th Large Hadron Collider Physics Conference (LHCP2024)\\
    3--7 June 2024 \\
    Boston, USA
}

\usepackage{xspace}

\newcommand{\fbinv}{\,\mbox{fb\textsuperscript{--1}}\xspace}
\newcommand{\TeV}{\,\mbox{Te\hspace{-.08em}V}\xspace}
\newcommand{\GeV}{\,\mbox{Ge\hspace{-.08em}V}\xspace}
\newcommand{\GeVns}{\mbox{Ge\hspace{-.08em}V}\xspace}

\begin{document}

\maketitle

\section{Introduction}

Heavy neutral leptons (HNLs) are introduced in proposed extensions of the standard model (SM) of particle physics to provide an explanation for the small but nonzero mass of the SM neutrinos via the see-saw mechanism.
HNLs can also be dark matter candidates, and can explain the matter-antimatter asymmetry in the universe via leptogenesis.
The recent theoretical and experimental progress in HNL physics is reviewed in Ref.~\cite{Antel:2023hkf}.

\begin{figure}[!ht]
\centering
\includegraphics[width=0.4\textwidth, page=1]{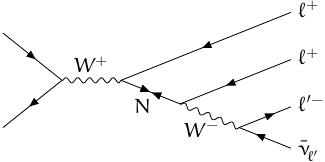}%
\hspace{0.1\textwidth}%
\includegraphics[width=0.4\textwidth, page=2]{feynman_hnl_trilep}
\caption{%
    Examples of leading-order Feynman diagrams for production and decay of an HNL (indicated with the symbol N) resulting in a final state with three charged leptons (left) or two charged leptons and two quarks (right).
    Here, the HNL is a Majorana particle, and thus the charged leptons from the HNL production and decay vertex can have the same charge.
}
\label{fig:feynman}
\end{figure}

At the CMS experiment~\cite{CMS:Detector-2008,CMS:Detector-2024}, various searches for HNL production in proton-proton (pp) collisions have been performed~\cite{CMS:EXO-23-006}.
The dominant production mode for HNLs with masses at the \GeVns scale is the Drell--Yan process where a W boson is produced and decays to an HNL and a charged lepton, as shown in Fig.~\ref{fig:feynman}.
For HNL masses above 20\GeV, the HNL decays promptly and its decay products will be associated with the primary vertex (PV) of the pp interaction.
In contrast, HNLs with masses below 20\GeV can be sufficiently long-lived such that the secondary vertex (SV) of the HNL decay can be reconstructed with the CMS tracker separately from the PV.

In this contribution, two HNL searches are presented.
The first search, published in Ref.~\cite{CMS:EXO-22-011}, considers promptly decaying HNLs in the mass range 10\GeV--1.5\TeV, where the fully leptonic HNL decay results in final states with three charged leptons all associated with the PV.
The second search, published in Ref.~\cite{CMS:EXO-21-011}, targets the displaced semileptonic decay of an HNL in the mass range 1--20\GeV, such that the signature is one charged lepton from the PV and a system of a charged lepton and a jet associated with an SV.
Both analyses are based on pp collisions data recorded at 13\TeV in 2016--2018, corresponding to an integrated luminosity of 138\fbinv, and employ machine-learning methods to separate the HNL signal from the SM background.

\section{Prompt search in trilepton final state}

\begin{figure}[!ht]
\vspace*{-\baselineskip}
\centering
\includegraphics[width=0.405\textwidth]{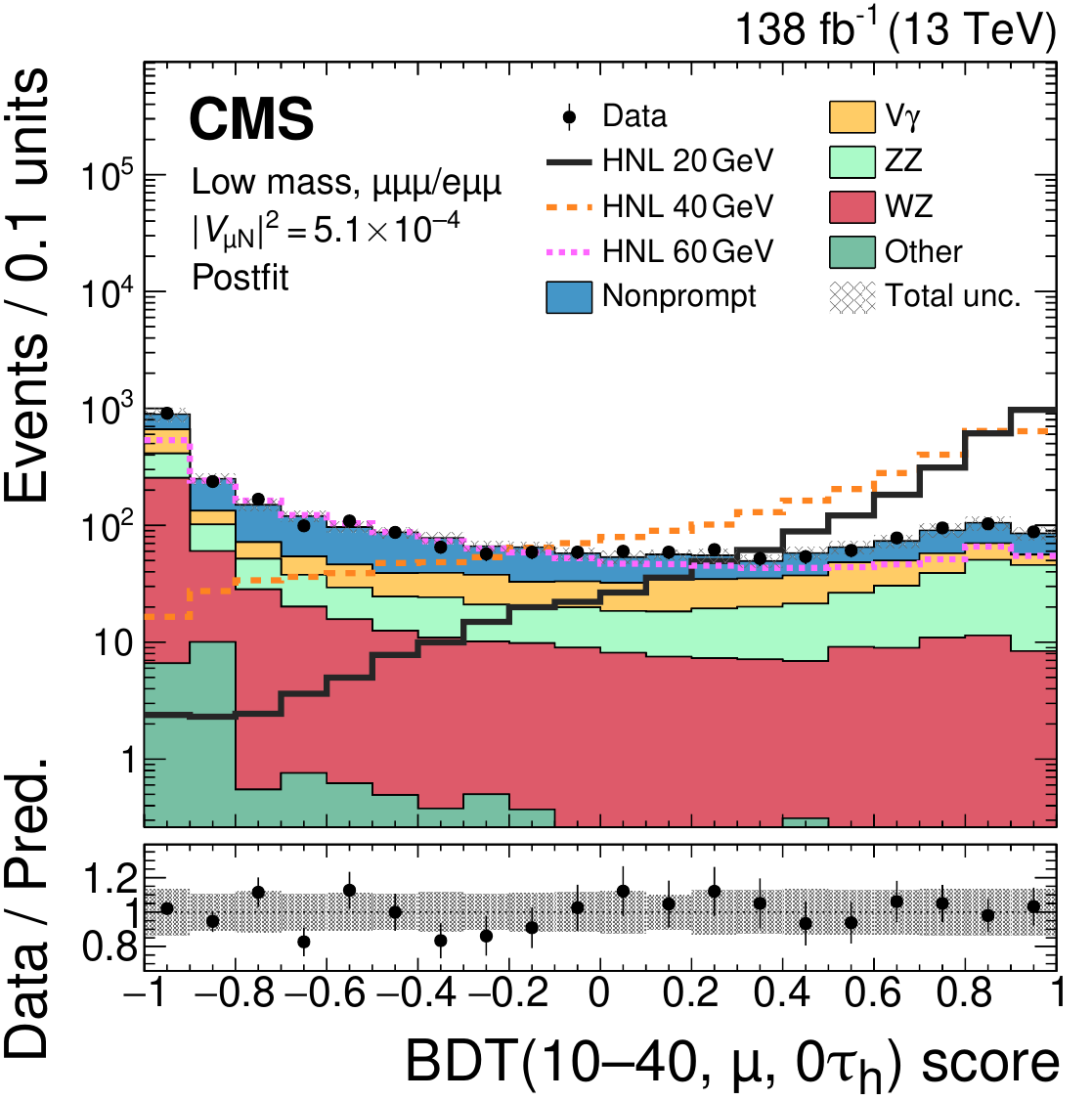}%
\hspace*{0.1\textwidth}%
\includegraphics[width=0.4\textwidth]{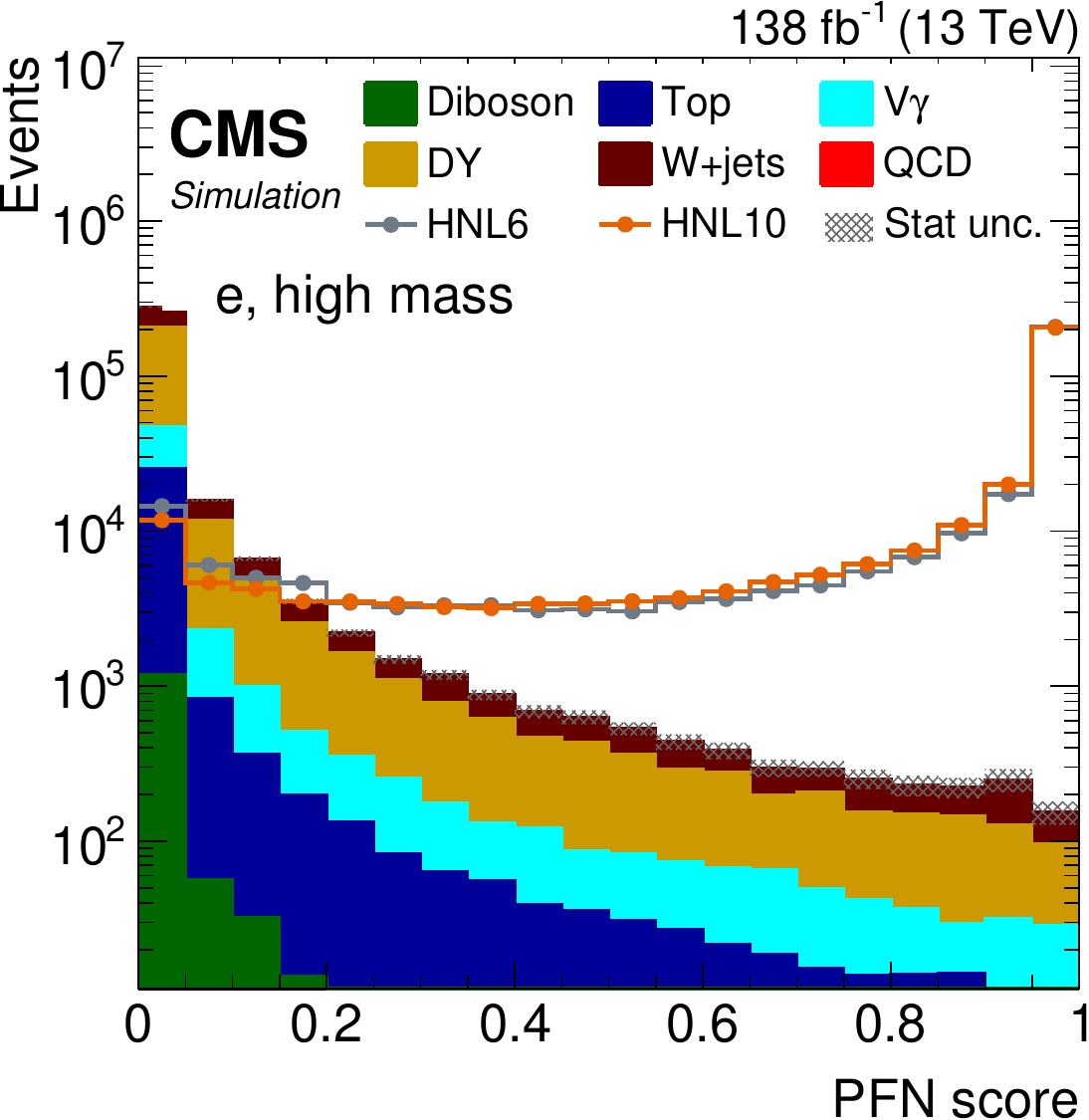}
\caption{%
    Left: Distribution of the BDT score trained with the 10--40\GeV HNL mass range and muon neutrino couplings for the combined \textmu\textmu\textmu\ and e\textmu\textmu\ categories of the prompt search in the trilepton final state, from Ref.~\cite{CMS:EXO-22-011}.
    Right: Distribution of the high-mass PFN score for the combined ee and \textmu e categories of the displaced search in the dilepton final state, from Ref.~\cite{CMS:EXO-21-011}.
}
\label{fig:discriminant}
\vspace*{-0.2\baselineskip}
\end{figure}

Events are selected with exactly three charged leptons from the PV, where up to one of them may be a reconstructed hadronically decaying tau lepton.
Boosted decision trees (BDTs) with up to 43 kinematic variables as input features are trained to separate the HNL signal from the SM background, separately for different HNL mass ranges, coupling scenarios, and final-state flavour combinations.
An example BDT score distribution is shown in Fig.~\ref{fig:discriminant} (left).
The BDT score distributions are used in maximum-likelihood fits to evaluate exclusion limits on the HNL coupling strength.

\section{Displaced search in dilepton final state}

Events are selected with exactly one electron or muon from the PV, and one additional nonprompt electron or muon associated with a jet and consistent with originating from an SV.
Particle flow networks (PFNs) with up to 50 input particles and additional event-level variables as input features are trained to distinguish whether the system of nonprompt lepton, jet, and SV originates from an HNL decay or from the SM background.
Separate PFNs are trained for different HNL mass ranges and both nonprompt lepton flavours.
An example PFN score distribution is shown in Fig.~\ref{fig:discriminant} (right).
The PFN score is used as one of two variables in the ABCD method to define a signal region enriched in signal-like events and to estimate the SM background from data in sideband regions.

\section{Results}

\begin{figure}[!ht]
\vspace*{-\baselineskip}
\centering
\includegraphics[width=0.45\textwidth]{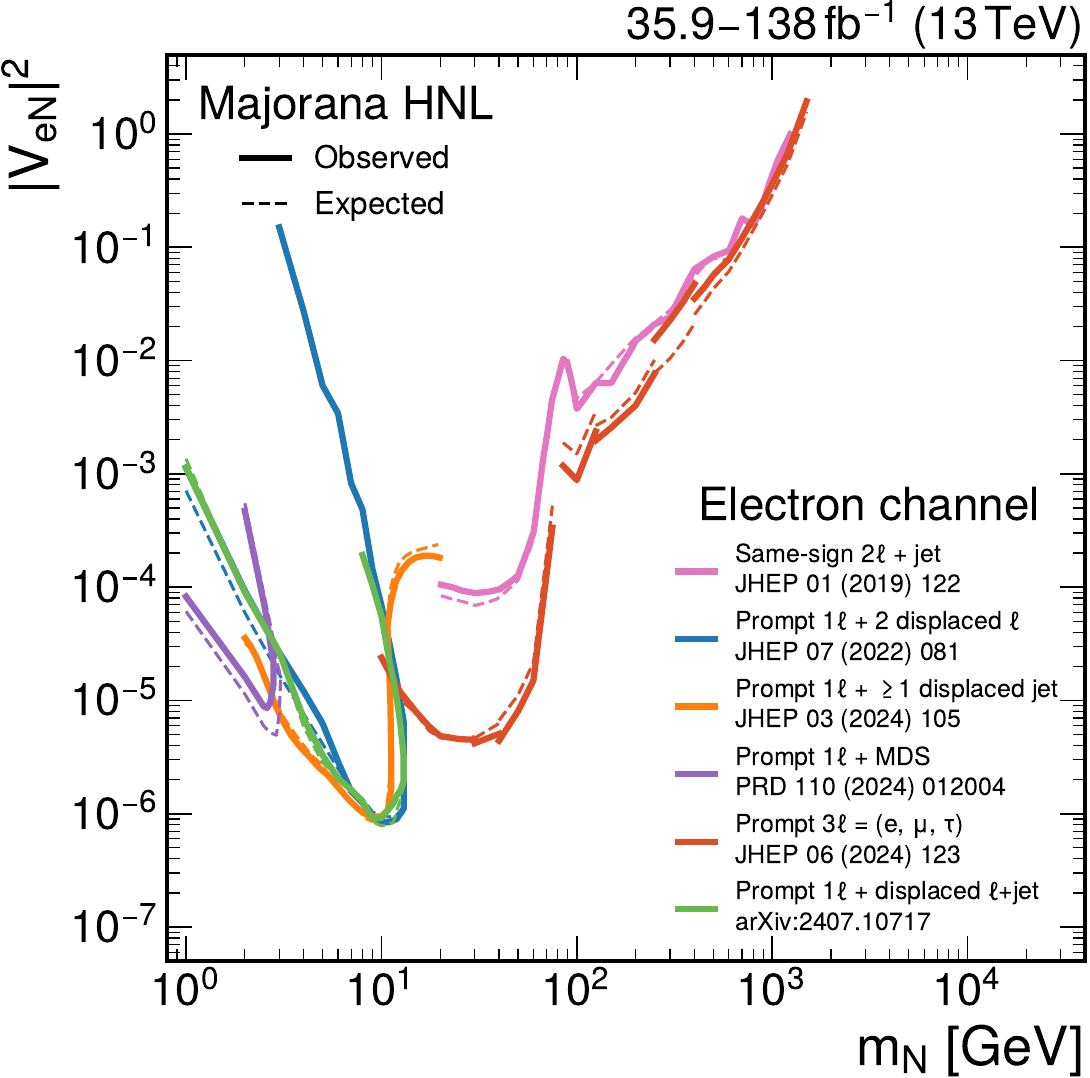}%
\hspace*{0.1\textwidth}%
\includegraphics[width=0.45\textwidth]{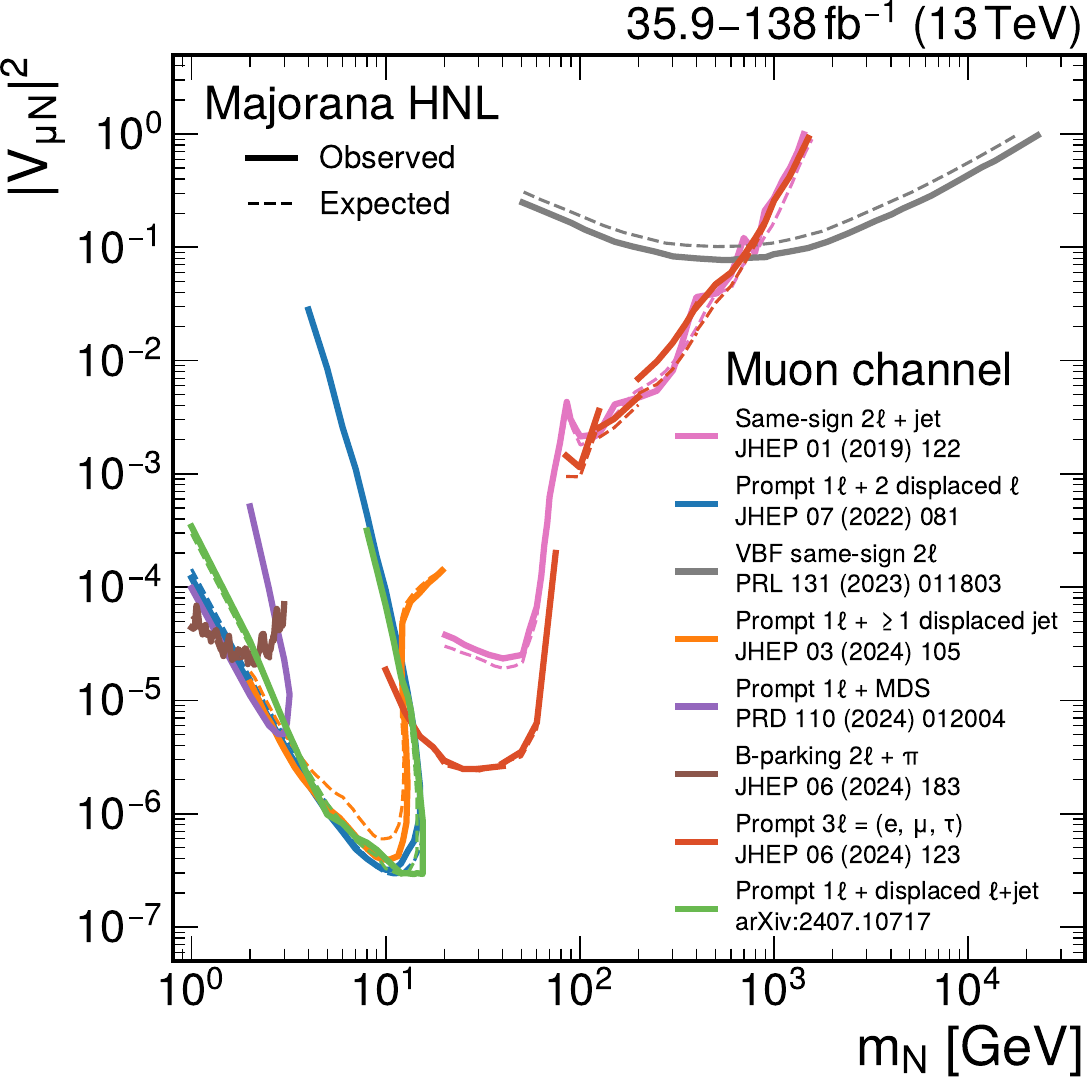} \\[1ex]
\includegraphics[width=0.45\textwidth]{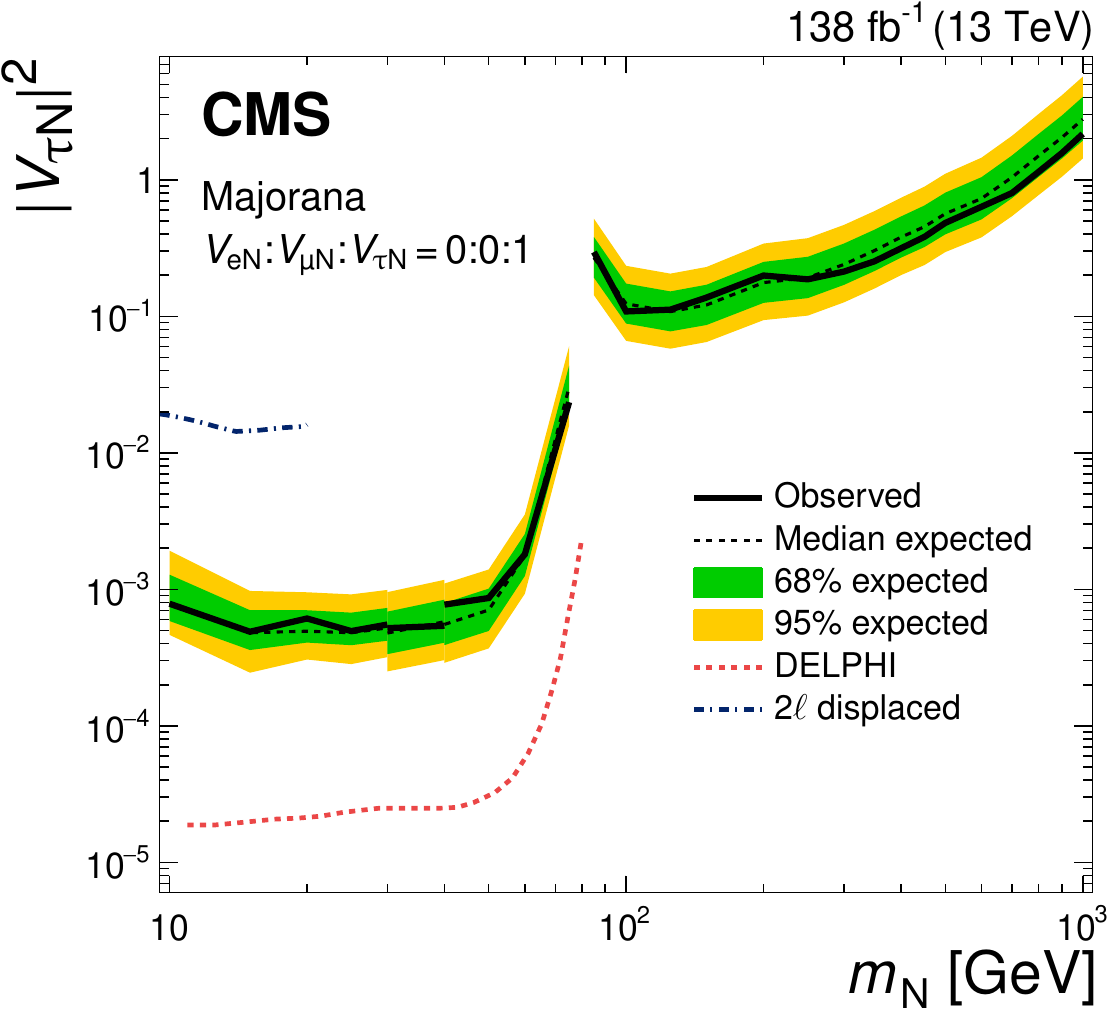}
\caption{%
    Exclusion limits at 95\% CL on the HNL mixing parameter as a function of the HNL mass for a Majorana HNL with exclusive electron (upper left), muon (upper right), or tau (lower, from Ref.~\cite{CMS:EXO-22-011}) neutrino coupling.
    The solid (dashed) curve indicates the observed (expected) exclusion.
}
\label{fig:limits}
\vspace*{-0.2\baselineskip}
\end{figure}

No significant deviations from the SM background are observed.
Exclusion limits on the HNL coupling strength as a function of the HNL mass are derived for HNLs of Dirac and Majorana nature and considering different coupling scenarios.
In Fig.~\ref{fig:limits}, the exclusion limits for the case of a Majorana HNL coupling exclusively to electron (upper left) or muon (upper right) neutrinos are shown, also compared to earlier CMS results in other channels~\cite{CMS:EXO-17-028, CMS:EXO-20-009, CMS:EXO-21-003, CMS:EXO-21-013, CMS:EXO-22-017, CMS:EXO-22-019}.
The prompt search in the trilepton final state~\cite{CMS:EXO-22-011} provides the most stringent limits over a wide mass range, and probes the exclusive couplings to tau neutrinos for HNL masses above the W boson mass for the first time, as shown in Fig.~\ref{fig:limits} (lower) and compared to an earlier DELPHI result~\cite{DELPHI:1996qcc}.
The displaced search in the dilepton final state~\cite{CMS:EXO-21-011} provides competitive limits with other analyses in the same mass range, and is particularly important for long-lived scenarios with HNL masses of 10--20\GeV.

\acknowledgments

The author acknowledges support from the Research Foundation Flanders (FWO) as a senior postdoctoral fellow fundamental research (grant number 1287324N).

\bibliographystyle{cms_unsrt_notitle}
\let\oldthebibliography\thebibliography
\renewcommand\thebibliography[1]{\oldthebibliography{#1}
    \setlength{\parskip}{0pt}
    \setlength{\itemsep}{0pt plus 0.3ex}
    \small
    \vspace*{-0.5\baselineskip}
}
\bibliography{refs.bib}

\end{document}